\documentclass[aps,prd,onecolumn,groupedaddress,showpacs,nofootinbib,amssymb]{revtex4}
\usepackage[dvips]{graphicx}
\usepackage{amssymb}
\usepackage{amsmath}
\usepackage{graphicx,,color}
\usepackage{amsfonts}
\usepackage{bm}
\usepackage{cancel}
\usepackage{comment}

\newcommand\be{\begin{equation}}
\newcommand\ee{\end{equation}}

\allowdisplaybreaks[4]

\begin{document}

\tolerance=5000

\title{GW170817-compatible Constant-roll Einstein-Gauss-Bonnet Inflation and Non-Gaussianities}
\author{S.D.~Odintsov,$^{1,2}$\,\thanks{odintsov@ieec.uab.es}
V.K.~Oikonomou,$^{3,4,5}$\,\thanks{v.k.oikonomou1979@gmail.com}F.P.
Fronimos,$^{3}$\,\thanks{fotisfronimos@gmail.com} S.A.
Venikoudis,$^{3}$\,\thanks{venikoudis@gmail.com}}
\affiliation{$^{1)}$ ICREA, Passeig Luis Companys, 23, 08010 Barcelona, Spain\\
$^{2)}$ Institute of Space Sciences (IEEC-CSIC) C. Can Magrans
s/n,
08193 Barcelona, Spain\\
$^{3)}$ Department of Physics, Aristotle University of
Thessaloniki, Thessaloniki 54124,
Greece\\
$^{4)}$ Laboratory for Theoretical Cosmology, Tomsk State
University of Control Systems and Radioelectronics, 634050 Tomsk,
Russia (TUSUR)\\
$^{5)}$ Tomsk State Pedagogical University, 634061 Tomsk,
Russia\\}

\tolerance=5000

\begin{abstract}
In this paper we investigate the inflationary phenomenology of an
Einstein-Gauss-Bonnet theory compatible with the GW170817 event,
by imposing the constant-roll evolution on the scalar field. We
develop the constant-roll GW170817-compatible
Einstein-Gauss-Bonnet formalism, and we calculate the slow-roll
indices and the observational indices of inflation, for several
models of interest. As we demonstrate, the phenomenological
viability of the models we study is achieved for a wide range of
the free parameters. In addition, for the same values of the free
parameters that guarantee the inflationary phenomenological
viability of the models, we also make predictions for the
non-Gaussianities of the models, since the constant-roll evolution
is known to enhance non-Gaussianities. As we show the
non-Gaussianities are of the same order for the slow-roll and
constant-roll case, and in fact in some cases, the amount of the
non-Gaussianities is smaller in the constant-roll case.
\end{abstract}


\maketitle

\section{Introduction}

After the striking neutron star merging GW170817 event
\cite{GBM:2017lvd}, which was followed by a kilonova, the fact
that the gravitational waves arrived almost simultaneously with
the electromagnetic radiation emitted by the kilonova, it was
obvious that the gravitational wave speed $c_T$ was nearly equal
to that of light's, that is $c_T^2=1$ in natural units. This fact
has put several generalizations of Einstein's theory of relativity
into peril, since several extended theories of gravity predict a
gravitational wave speed different from that of light's, referring
always to the speed of the tensor perturbations. One question with
conceptual interest is, does it matter if some theories of
extended gravity predict a primordial gravitational wave speed
different than that of light's? The answer could be easy to answer
by simply thinking that the Universe during the inflationary era
and in the post-inflationary era, is classical, and described by a
four dimensional spacetime metric. Thus there is no particle
physics reason for the Universe to change the mass of the
primordial graviton. Thus indeed, the graviton, either it
propagates in the form of primordial gravitational waves, or it
propagates in the form of astrophysical originating gravitational
waves, should be massless, or nearly massless. An extensive list
of theories which were put into question after the GW170817 event,
can be found in Ref. \cite{Ezquiaga:2017ekz}.

One of the theories that were put into question after the GW170817
event, were the Einstein-Gauss-Bonnet theories
\cite{Hwang:2005hb,Nojiri:2006je,Cognola:2006sp,Nojiri:2005vv,Nojiri:2005jg,Satoh:2007gn,Bamba:2014zoa,Yi:2018gse,Guo:2009uk,Guo:2010jr,Jiang:2013gza,Kanti:2015pda,vandeBruck:2017voa,Kanti:1998jd,Pozdeeva:2020apf,Fomin:2020hfh,DeLaurentis:2015fea,Chervon:2019sey,Nozari:2017rta,Odintsov:2018zhw,Kawai:1998ab,Yi:2018dhl,vandeBruck:2016xvt,Kleihaus:2019rbg,Bakopoulos:2019tvc,Maeda:2011zn,Bakopoulos:2020dfg,Ai:2020peo,Odintsov:2019clh,Oikonomou:2020oil,Odintsov:2020xji,Oikonomou:2020sij,Odintsov:2020zkl,Odintsov:2020sqy,Easther:1996yd,Antoniadis:1993jc,Antoniadis:1990uu,Kanti:1995vq,Kanti:1997br,Bajardi:2019zzs,Capozziello:2019wfi,deMartino:2020yhq},
see also the review \cite{Nojiri:2010wj} which form an appealing
class of theories capable of describing the inflationary era and
also several astrophysical objects. The reason for considering
Einstein-Gauss-Bonnet theories as appealing candidate theories for
the primordial era of our Universe is simply because these are
string motivated theories, basically the whole theory is a
string-corrected canonical scalar field theory minimally coupled
to gravity. In several previous works
\cite{Odintsov:2019clh,Oikonomou:2020oil,Odintsov:2020xji,Odintsov:2020zkl,Odintsov:2020sqy,Oikonomou:2020sij}
we demonstrated how Einstein-Gauss-Bonnet theories and their
extensions, may actually be rectified in view of the GW170817
event, by simply demanding that the primordial gravitational wave
speed is set equal to unity. In effect, this constraint results to
a differential equation which constrains severely the functional
form of the scalar potential $V(\phi)$ and of the scalar coupling
function $\xi(\phi)$ of the scalar field with the Gauss-Bonnet
invariant.

In this paper, we shall extend the formalism of our previous work
\cite{Odintsov:2020sqy}, to take into account a constant-roll
evolution for the scalar field. The constant-roll evolution is a
widely used assumption for the evolution of the scalar field
during the primordial era. The aim of this paper is two-fold:
Firstly we shall investigate whether a viable phenomenology can be
obtained by the constant-roll GW170817-compatible
Einstein-Gauss-Bonnet theory. Secondly, we shall investigate what
is the predicted amount of non-Gaussianities predicted by the
GW170817-compatible Einstein-Gauss-Bonnet theory, when the
constant-roll assumption is used, since the constant-roll
evolution is known to enhance the non-Gaussianities features. Our
results are quite interesting, since we evince that the
constant-roll evolution assumption for the scalar field can also
yield the GW170817-compatible Einstein-Gauss-Bonnet theory viable
and very good aligned with the latest Planck data
\cite{Akrami:2018odb}, but more importantly, the non-Gaussianities
in the case at hand are not enhanced, and in some cases are
smaller in value, when compared to the slowly rolling scalar field
scenario for the GW170817-compatible Einstein-Gauss-Bonnet theory.
Our motivation to use modified gravity description for the
inflationary era, comes from the fact that general relativity
seems to fail to consistently describe several evolution eras of
our Universe, such as the dark energy era, and in some cases the
inflationary era, see for reviews
\cite{Nojiri:2017ncd,Capozziello:2011et,Capozziello:2010zz,Nojiri:2006ri,
Nojiri:2010wj,delaCruzDombriz:2012xy,Olmo:2011uz}. Also in some
cases, it is possible that modified gravity can mimic dark matter,
but also dark matter can also be a massive particle with no
interaction or small interaction with other particles
\cite{Bertone:2004pz,Bergstrom:2000pn,Mambrini:2015sia,Profumo:2013yn,Hooper:2007qk,Oikonomou:2006mh}.

Before starting, an important discussion is in order. The Planck
2018 data on inflation are able to bring information relevant to
the inflationary era, available and unaltered at late-times, due
to the mechanism of inflation itself. Basically, the information
measured in the CMB at present, is nothing else but the primordial
modes which exited the Hubble horizon at the first time, at the
time instance we assumed that inflation started. These modes were
frozen after the horizon crossing, and re-entered the Hubble
horizon during the radiation and matter domination eras,
unaltered. For the latter reason the primordial modes carry
information about the inflationary era, these are the frozen modes
at early times. Now regarding the primordial tensor modes, the
same principle applies, hence if primordial gravitational waves
are ever found, these must be massless modes and which correspond
to a gravitational wave speed equal to unity. Now the question is
whether someone should expect these primordial modes to be
massless, and why should an astrophysical gravitational wave speed
equal to unity, impose constraints on the early-time primordial
gravity waves. From a fundamental physics point of view, gravity
is mediated by gravitons, so regardless the graviton mediates
primordial gravity waves, or astrophysical gravity waves, the
graviton is the same. From a particle physics point of view, there
is no fundamental reason for the graviton to alter its mass during
the inflationary and the post-inflationary era. This is why the
constraint brought along by the kilonova related event GW170817
for a massless graviton, also affects the early-time tensor
perturbation modes, thus the primordial gravity wave speed. For
us, Einstein-Gauss-Bonnet theory is one of the most appealing
extensions of minimally coupled scalar field theory, since it is
string motivated and also with our approach, the gravity waves
which are basically the primordial tensor modes, are also
massless, as the present time graviton seems to be.

\section{Constant-roll Inflationary Evolution of Einstein-Gauss-Bonnet Gravity}

In this section we shall investigate how the theoretical framework
of Ref. \cite{Odintsov:2020sqy} is modified if the constant-roll
evolution is adopted for the scalar field. In order to render the
article self-contained, we shall describe in brief the formalism
of the GW170817-compatible Einstein-Gauss-Bonnet gravity developed
in Ref. \cite{Odintsov:2020sqy}, and we shall consider a minimally
coupled Einstein Gauss-Bonnet theory described by the
gravitational action,
\begin{equation}
\centering
\label{action}
S=\int {d^4x\sqrt{-g}\left( \frac{R}{2\kappa^2}-\frac{1}{2}\omega g^{\mu\nu}\partial_\mu\phi\partial_\nu\phi-V(\phi)-\xi(\phi)\mathcal{G}\right)}\,
\end{equation}
where $g$ is the metric determinant, $R$ denotes the Ricci scalar,
$\kappa=\frac{1}{M_P}$ is the gravitational constant where $M_P$
denotes the reduced Planck mass, and $V(\phi)$ is the scalar
potential, while $\mathcal{G}$ describes the Gauss-Bonnet
invariant
$\mathcal{G}=R^2-4R_{\alpha\beta}R^{\alpha\beta}+R_{\alpha\beta\gamma\delta}R^{\alpha\beta\gamma\delta}$,
with $R_{\alpha\beta}$ and $R_{\alpha\beta\gamma\delta}$ being the
Ricci and Riemann tensor respectively. Finally, $\xi(\phi)$
denotes the Gauss-Bonnet coupling scalar function. Moreover, we
shall assume that the geometric background is a flat
Friedman-Robertson-Walker background, with the line element being,
\begin{equation}
\centering
\label{metric}
ds^2=-dt^2+a(t)^2\delta_{ij}dx^idx^j.\,
\end{equation}
According to this form, the metric tensor reads
$g_{\mu\nu}=diag(-1, a(t)^2, a(t)^2, a(t)^2)$. Furthermore, we
shall also assume that the scalar field $\phi$ is homogeneous, or
in other words it is only time-dependent. Furthermore, since the
metric is flat, the Ricci scalar and the Gauss-Bonnet invariant
can be written in very simple forms, as $R=12H^2+6\dot H$ and
$\mathcal{G}=24H^2(\dot H+H^2)$. Here, $H$ signifies Hubble's
parameter and in addition, the ``dot'' denotes differentiation
with respect to the cosmic time as usual. Finally, we note that
the term $\omega$ in the kinetic term will be set equal to unity
in order to describe the canonical case, but for the time being we
shall leave it as it is in order to show how the results depend on
such term. However it shall be treated as a constant, independent
of the scalar field.

By varying the gravitational action (\ref{action}), one can
extract the field equations easily. Consequently, the equations of
motion are derived easily from the time and space components of
the field equations for gravity and the continuity equation of the
scalar field, which read,
\begin{equation}
\centering
\label{motion1}
\frac{3H^2}{\kappa^2}=\frac{1}{2}\omega\dot\phi^2+V+24\dot\xi H^3,\,
\end{equation}
\begin{equation}
\centering
\label{motion2}
\frac{-2\dot H}{\kappa^2}=\omega\dot\phi^2-16\dot\xi H\dot H+8H^2(\ddot\xi-H\dot\xi),\,
\end{equation}
\begin{equation}
\centering
\label{motion3}
\ddot\phi+3H\dot\phi+\frac{1}{\omega}\left(V'+\xi'\mathcal{G}\right)=0,\,
\end{equation}
where in contrast to the previous notation, the prime denotes
differentiation with respect to the scalar field $\phi$.
Describing the inflationary era properly implies an analytical
solution of the system of equations of motion. Unfortunately, such
a system is very difficult to study analytically. The solution can
however be extracted by assuming certain approximations during
inflation, after the first horizon crossing, or the initial moment
of inflation. Here, we shall assume that the slow-roll
approximations hold true and also impose the constant-roll
condition on the scalar field. Mathematically speaking, we shall
assume that the following conditions hold true,
\begin{align}
\label{approx}
\centering
\dot H&\ll H^2,& \frac{1}{2}\omega\dot\phi^2&\ll V,& \ddot\phi=\beta H\dot\phi,\,
\end{align}
where $\beta$ is the constant-roll parameter. These assumptions
make the equations of motion simpler and we end up with the
following expressions,
\begin{equation}
\centering
\label{motion4a}
H^2\simeq\frac{\kappa^2V}{3},\,
\end{equation}
\begin{equation}
\label{motion5a}
\centering
\dot H\simeq-\frac{1}{2}\kappa^2\omega\dot\phi^2,\,
\end{equation}
\begin{equation}
\centering
\label{motion6a}
V'+(3+\beta)\omega H\dot\phi+24\xi'H^4\simeq0.\,
\end{equation}
These are the simplified equations of motion we shall use in order
to produce results. However, before we proceed further, we shall
impose certain additional constraints in order to achieve
compatibility with recent striking observations.

The tensor perturbations of the flat FRW metric, or simply the
primordial gravitational waves as they are called, propagate
through spacetime with the velocity of light, as it was recently
ascertained by the GW170817 event. This realization made it
abundantly clear that theories which describe modified gravity and
produce a different velocity must be discarded. A theory which
belongs to that category is the Einstein-Gauss-Bonnet theory,
since string corrections produce the following expression for
their velocity in natural units,
\begin{equation}
\centering
c_T^2=1-\frac{Q_f}{2Q_t},\,
\end{equation}
where $Q_f= 16(\ddot\xi-H\dot\xi)$ and
$Q_t=\frac{1}{\kappa^2}-8\dot\xi H$. The compatibility with the
GW170817 event may be achieved only if we demand $c_T^2=1$. This
in turn implies that the numerator of the second term becomes
zero, or in other words $\ddot\xi=H\dot\xi$. This is an ordinary
differential equation which can be solved easily. Although finding
an expression for the term $\dot\xi(\phi)$, which satisfies the
aforementioned differential equation, is feasible
\cite{Odintsov:2020sqy}, we shall choose a different approach
taking advantage of the constant-roll condition. Let us expand the
differential equation with respect to the scalar field. Since
(\ref{approx}) holds true, and the differential operator
$\frac{d}{dt}$ is equivalent to $\dot\phi\frac{d}{d\phi}$, then we
have,
\begin{equation}
\centering
\xi''\dot\phi^2+\beta H\xi'\dot\phi=H\xi'\dot\phi.\,
\end{equation}
Therefore, the expression for the derivative of the scalar field
is,
\begin{equation}
\label{dotphi}
\centering
\dot\phi=(1-\beta)H\frac{\xi'}{\xi''}.\,
\end{equation}
Taking the limit $\beta=0$, the above formula in equivalent to the
analysis made in a previous work of ours \cite{Odintsov:2020sqy}
where we studied the slow-roll case, as expected. Thus, the
equations of motion are rewritten in the case at hand as follows,
\begin{equation}
\centering
\label{motion4}
H^2\simeq\frac{\kappa^2V}{3},\,
\end{equation}
\begin{equation}
\label{motion5}
\centering
\dot H\simeq-\frac{H^2}{2}\kappa^2\omega(1-\beta)^2\left(\frac{\xi'}{\xi''}\right)^2,\,
\end{equation}
\begin{equation}
\label{motion6}
\centering
V'+(1-\beta)(1+\frac{\beta}{3})\kappa^2\omega\frac{\xi'}{\xi''}V+\frac{8}{3}\kappa^4\xi'V^2\simeq0.\,
\end{equation}
The above set of equations is much more easy to manipulate
analytically, as we show in the next sections. Firstly, Eq.
(\ref{motion5}) is connected to the slow-roll index $\epsilon_1$
as we shall see in the subsequent calculations. It is a useful
expression since it is interconnected to the constant-roll
parameter $\beta$ and the ratio of derivatives of the Gauss-Bonnet
coupling function. Hence, designating an appropriate coupling
function is of fundamental importance. Furthermore, the degrees of
freedom have decreased by one, since constraints on the velocity
of the gravitational waves were imposed. In consequence,
specifying the coupling function leads to a differential equation
which, once it is solved, it generates the scalar potential.
Hence, these terms cannot be designated freely but they are in
fact interconnected, although they have different origins. Let us
now proceed with the evaluation of the slow-roll indices.

The dynamics of inflation can be described by six parameters named
the slow-roll indices, defined as follows
\cite{Hwang:2005hb,Odintsov:2020sqy},
\begin{align}
\centering \epsilon_1&=-\frac{\dot
H}{H^2},&\epsilon_2&=\frac{\ddot\phi}{H\dot\phi},&\epsilon_3&=\frac{\dot
F}{2HF},&\epsilon_4&=\frac{\dot E}{2HE},&\epsilon_5&=\frac{\dot
F+Q_a}{2HQ_t},&\epsilon_6&=\frac{\dot Q_t}{2HQ_t},\,
\end{align}
where $F=\frac{1}{\kappa^2}$, $Q_a=-8\dot\xi H^2$,
$Q_t=\frac{1}{\kappa^2}-8\dot\xi H$ and
$E=\frac{1}{\kappa^2}+\frac{3Q_a^2}{2\dot\phi^2Q_t}$. Hence,
according to equations (\ref{approx}), (\ref{dotphi}),
(\ref{motion5}) and (\ref{motion6}), the indices can be rewritten
as follows,
\begin{equation}
\centering
\label{index1}
\epsilon_1\simeq\frac{\kappa^2\omega}{2}(1-\beta)^2\left(\frac{\xi'}{\xi''}\right)^2,\,
\end{equation}
\begin{equation}
\label{index2}
\epsilon_2=\beta,\,
\end{equation}
\begin{equation}
\label{index3}
\centering
\epsilon_3=0,\,
\end{equation}
\begin{equation}
\label{index4}
\centering
\epsilon_4=\frac{1-\beta}{2}\frac{\xi'}{\xi''}\frac{E'}{E},\,
\end{equation}
\begin{equation}
\label{index5}
\centering
\epsilon_5\simeq-\frac{4(1-\beta)\xi'^2\kappa^4V}{3\xi''-8(1-\beta)\xi'^2\kappa^4V},\,
\end{equation}
\begin{equation}
\label{index6}
\centering
\epsilon_6\simeq-\frac{4(1-\beta)\xi'^2\kappa^4V(1-\epsilon_1)}{3\xi''-8(1-\beta)\xi'^2\kappa^4V}.\,
\end{equation}
Obviously, in the limit $\beta=0$, all the slow-roll indices,
apart from $\epsilon_2$, are restored as in the slow-roll case we
studied in Ref. \cite{Odintsov:2020sqy}. Moreover, it is clear
that the value $\beta=1$, which in turn implies that
$\ddot\phi=H\dot\phi$, is not an accepted value for $\beta$ due to
the fact that it leads to zero slow-roll indices, apart from
$\epsilon_2$, and this choice would lead to eternal inflation.
This was obviously implied previously when we performed a division
with $\dot\phi$ in order to extract the form of $\dot\phi$
depending on the coupling scalar function and Hubble's parameter.
On the other hand, there exists no physical constraint which
prohibits $\beta$ to obtain values greater than unity, so the
inequality $\beta>1$ could still yield interesting phenomenology.
We also mention that the auxiliary functions are written as,
\begin{equation}
\centering
E=\frac{1}{\kappa^2}+\frac{96}{\kappa^2Q_t}\xi'^2H^4,\,
\end{equation}
\begin{equation}
\centering
Q_a=-8(1-\beta)\frac{\xi'^2}{\xi''}H^3,\,
\end{equation}
\begin{equation}
\centering
Q_t=\frac{1}{\kappa^2}-8(1-\beta)\frac{\xi'^2}{\xi''}H^2,\,
\end{equation}
\begin{equation}
\centering
Q_e=-32(1-\beta)\frac{\xi'^2}{\xi''}H\dot H.\,
\end{equation}
The term $Q_e$ was introduced here, but will be used in the
following relations. Lastly, we discuss the form of the
observational indices in the case of the model at hand. The
spectral index of primordial curvature perturbations $n_S$, the
spectral index of tensor perturbations $n_T$ and the
tensor-to-scalar ratio $r$ in terms of the slow-roll indices, are
defined as follows \cite{Hwang:2005hb,Odintsov:2020sqy},
\begin{align}
\label{observed}
\centering
n_s&=1-2\frac{2\epsilon_1+\epsilon_2+\epsilon_4}{1-\epsilon_1},&n_T&=-2\frac{\epsilon_1+\epsilon_6}{1-\epsilon_1},&r&=16\left|\left(\frac{\kappa^2Q_e}{4H}-\epsilon_1\right)\frac{c_A^3}{\kappa^2Q_t}\right|,\,
\end{align}
where  $c_A$ the sound wave velocity defined as,
\begin{equation}
\centering
\label{soundwave}
c_A^2=1+\frac{Q_aQ_e}{3Q_a^2+2Q_t\omega\dot\phi^2}.\,
\end{equation}
The aim in the rest of the paper is to evaluate the observational
indices during the first horizon crossing. However, instead of
using wavenumbers, we shall use the values of the scalar potential
during the initial stage of inflation. Taking it as an input, we
can obtain the actual values of the observational quantities. We
can do so by firstly evaluating the final value of the scalar
field. This value can be derived by equating slow-roll index
$\epsilon_1$ in Eq. (\ref{index1}) to unity. Consequently, the
initial value can be evaluated from the $e$-foldings number,
defined as
$N=\int_{t_i}^{t_f}{Hdt}=\int_{\phi_i}^{\phi_f}{\frac{H}{\dot\phi}d\phi}$,
where the difference $t_f-t_i$ signifies the duration of the
inflationary era. Recalling the definition of $\dot\phi$ in Eq.
(\ref{dotphi}), one finds that the proper relation from which the
initial value of the scalar field can be derived is,
\begin{equation}
\centering
\label{efolds}
N=\frac{1}{1-\beta}\int_{\phi_i}^{\phi_f}{\frac{\xi''}{\xi'}d\phi}.\,
\end{equation}
From this equation, as well as equation (\ref{index1}), it is
obvious that choosing an appropriate coupling function, is the key
in order to simplify the results. In the following, we shall work
with certain functional forms of this coupling function, derive
the scalar potential from (\ref{motion6}) and produce results for
both the observational quantities introduced previously, but also
we shall discuss the primordial non-Gaussianities, known to occur
when the constant-roll condition is used, as we mentioned in the
introduction. In the following section, we shall introduce the
formalism of non-Gaussianities before continuing with examining
the viability of certain models.

\section{Primordial Non Gaussianities Under The Constant-Roll Condition}

Until now, the perturbations in the Cosmic Microwave Background
(CMB) are described perfectly as Gaussian distributions, since no
practical evidence is found pointing out a non-Gaussian pattern in
the CMB. It is possible though, not probable for the moment, that
in the following years the observations may reveal a non-Gaussian
pattern in the CMB primordial power spectrum. In this section we
shall discuss how to evaluate the non-Gaussianities quantitatively
in the context of the GW170817-compatible Einstein-Gauss-Bonnet
gravity, using the formalism and notation of
\cite{DeFelice:2011zh}. We first define the following quantities,
\begin{align}
\centering
\delta_\xi&=\kappa^2H\dot\xi,&\delta_X&=\frac{\kappa^2\omega\dot\phi^2}{H^2},&\epsilon_s&=\epsilon_1-4\delta_\xi,&n&=\frac{\dot\epsilon_s}{H\epsilon_s},&s&=\frac{\dot
c_A}{Hc_A}.\,
\end{align}
Here, we shall implement a different formula for the sound wave
speed, however equivalent to the previous, which is based on these
newly defined quantities for convenience and reads,
\begin{equation}
\centering
c_A^2\simeq1-\frac{64\delta_\xi^2(6\delta_\xi+\delta_X)}{\delta_X}.\,
\end{equation}
Recalling equations (\ref{dotphi}), (\ref{motion5}) and
(\ref{motion6}), one finds that the previous auxiliary terms have
the following forms,
\begin{equation}
\centering
\delta_\xi\simeq\frac{1-\beta}{3}\kappa^4V\frac{\xi'^2}{\xi''},\,
\end{equation}
\begin{equation}
\centering
\delta_X\simeq\kappa^2\omega(1-\beta)^2\left(\frac{\xi'}{\xi''}\right) ^2=2\epsilon_1,\,
\end{equation}
\begin{equation}
\centering
\epsilon_s\simeq(1-\beta)\left(\frac{\kappa\xi'}{\xi''}\right)^2\left(\frac{\omega(1-\beta)}{2}-\frac{4}{3}\kappa^2\xi''V\right),\,
\end{equation}
\begin{equation}
\centering
n\simeq2(1-\beta)\left(1-\frac{\xi'\xi'''}{\xi''^2}-4\kappa^2\frac{\xi'}{\xi''}\frac{V'\xi''+V\xi'''}{3\omega(1-\beta)-8\kappa^2\xi''V}\right),\,
\end{equation}
\begin{equation}
\centering
s=(1-\beta)\frac{\xi'}{\xi''}\frac{c_A'}{c_A}.\,
\end{equation}
In certain examples, we shall see that by choosing appropriately
the coupling function simplifies greatly the quantity
$\epsilon_s$, as it shall also coincide with $\epsilon_1$, along
with $\delta_X$. No matter the form of the sound wave velocity,
the derivative $c_A$ is very complex, so we omit its analytic
expression. These forms are very useful due to the fact that the
power spectra $\mathcal{P}_S$ of the primordial curvature
perturbations and the equilateral momentum approximation term
$f_{NL}^{eq}$ can be derived from such terms. These quantities are
defined as,
\begin{equation}
\centering
\label{spectra}
\mathcal{P}_S=\frac{\kappa^4V}{24\pi^2\epsilon_sc_A},\,
\end{equation}
\begin{equation}
\centering
\label{NL}
f_{NL}^{eq}\simeq\frac{55}{36}\epsilon_s+\frac{5}{12}n+\frac{10}{3}\delta_\xi.\,
\end{equation}
In the following we shall appropriately specify the value of the
term $f_{NL}^{eq}$ during the first horizon crossing, to see what
the constant-roll condition brings along. The evaluation shall be
performed by using the values of the free parameters in such a way
so that the viability of the observational indices of inflation
are compatible with the 2018 Planck data \cite{Akrami:2018odb}.

It is useful to note that, the spectral indices of scalar and
tensor perturbations and the tensor-to-scalar ratio which will be
numerically evaluated in the subsequent sections for appropriately
chosen models, can be derived using the auxiliary parameters of
this section, as follows,
\begin{align}
\centering
n_S&=1-2\epsilon_s-n-s-8\delta_\xi,&n_T&=-2\epsilon_s-8\delta_\xi,&r&=16\frac{\epsilon_sc_A}{1-8\delta_\xi}.\,
\end{align}
These are obviously equivalent to the definitions presented in the
previous section, but we shall proceed with the slow-roll
expression.

\section{Specific Models and Their Compatibility with Recent Observations}

As it was mentioned before, our main aim is to extract the value
of the scalar field during the first horizon crossing and insert
it as an input in Eq. (\ref{observed}). Firstly, we shall define
the Gauss-Bonnet coupling scalar function. Afterwards, we shall
derive the scalar potential from Eq. (\ref{motion6}) corresponding
to the selected coupling function. Accordingly, we shall equate
the slow-roll index $\epsilon_1$ (\ref{index1}) with unity in
order to find the final value of the scalar field and finally,
from Eq. (\ref{efolds}) the initial value of the scalar field will
be extracted.

Let us now discuss several models which can produce viable
results.

\subsection{Model I: Power-Law Coupling Function}

Suppose that the Gauss-Bonnet coupling scalar function is defined
as follows,
\begin{equation}
\centering
\label{xi1}
\xi(\phi)=\lambda_1(\kappa\phi)^{m_1},\,
\end{equation}
where $\lambda_1$ is an unspecified for the time being
dimensionless constant. This is a very appealing function since
the ratio $\xi'\xi''$ which appears in our calculations is greatly
simplified, since,
\begin{equation}
\centering
\xi''=\frac{m_1-1}{\phi}\xi'.\,
\end{equation}
This is a model which was also studied in Ref.
\cite{Odintsov:2020sqy}. Since the coupling function is specified,
the scalar potential can be derived from Eq. (\ref{motion6}). The
form of the potential is very intricate as shown below,
\begin{equation}
\centering
\label{potA}
V(\phi)=\frac{3 e^{-\alpha_1(\kappa\phi)^2} \left(6\alpha_1 (\kappa\phi)^2\right)^{m_1/2}}{3 c \left(6\alpha_1 (\kappa\phi)^2 \right)^{m_1/2}-4\lambda_1  6^{\frac{m_1}{2}} m_1\kappa^4 (\kappa  \phi )^{m_1} \Gamma \left(\frac{m_1}{2},\alpha_1 (\kappa\phi)^2\right)},\,
\end{equation}
where $\alpha_1=-\frac{(\beta^2+2\beta-3)\omega}{6(m_1-1)}$ and
$c$ is the integration constant with mass dimensions [m]$^{-4}$,
and $\Gamma\left(\frac{m_1}{2},\alpha_1 (\kappa\phi)^2\right)$ is
the incomplete from below gamma function. Let us now proceed with
the evaluation of the slow-roll indices and certain auxiliary
parameters. These are,
\begin{equation}
\centering
\label{dxiA}
\delta_\xi\simeq\frac{(1-\beta ) \lambda_1  m_1 \kappa^4V(\phi ) (\kappa  \phi )^m_1}{3 (m_1-1)},\,
\end{equation}
\begin{equation}
\centering
\label{esA}
\epsilon_s\simeq\frac{(1-\beta ) \left(3 (1-\beta )(\kappa \phi) ^2 \omega -8 \lambda_1  (m_1-1) m_1 \kappa^4V(\phi ) (\kappa  \phi )^{m_1}\right)}{6 (m_1-1)^2},\,
\end{equation}
\begin{equation}
\centering
\label{index1A}
\epsilon_1\simeq\frac{\omega}{2}\left(\frac{1-\beta}{m_1-1}\right)^2(\kappa\phi)^2,\,
\end{equation}
\begin{equation}
\centering
\label{index2A}
\epsilon_2=\beta,\,
\end{equation}
\begin{equation}
\centering
\label{index3A}
\epsilon_3=0,\,
\end{equation}
\begin{equation}
\label{index5A}
\centering
\epsilon_5\simeq\frac{4 (1-\beta ) \lambda_1  m_1 \kappa^4V(\phi ) (\kappa  \phi )^{m_1}}{8 (1-\beta ) \lambda_1  m_1 \kappa^4V(\phi ) (\kappa  \phi )^{m_1}-3 (m_1-1)},\,
\end{equation}
\begin{equation}
\centering
\label{index6A}
\epsilon_6\simeq\frac{4 (1-\beta)^2  \lambda  m_1 (\kappa  \phi )^{m_1} \left(m_1 \kappa^4V(\phi )+\kappa\phi  \kappa^3V'(\phi )\right)}{(m_1-1) \left(8 (1-\beta )  \lambda_1  m_1 \kappa^4V(\phi ) (\kappa  \phi )^{m_1}-3 (m_1-1)\right)}.\,
\end{equation}
It is obvious that only the first three slow-roll indices have
simple forms while the rest have very perplexed, since they depend
on the scalar potential presented previously. This is exactly why
the index $\epsilon_4$ was not written analytically. Due to the
simple expression of index $\epsilon_1$ however, we can evaluate
the final value of the scalar field during the inflationary era by
equating $\epsilon_1$ with unity. Therefore, the resulting form
is,
\begin{equation}
\centering
\label{scalarfA}
\phi_f=\pm\frac{1}{\kappa}\sqrt{\frac{2}{\omega}}\left|\frac{m_1-1}{1-\beta}\right|.\,
\end{equation}
As a result, the initial value of the scalar field, which is also
the one that we need in order to evaluate both the observational
quantities and the predicted non-Gaussianities of the model, can
be extracted directly from equation (\ref{efolds}). The resulting
value is,
\begin{equation}
\centering
\label{scalariA}
\phi_i=\phi_fe^{-\frac{N(1-\beta)}{m_1-1}}.\,
\end{equation}
In this case, we shall use the positive values of the scalar
field. Assuming that in reduced Planck Units, where $\kappa^2=1$,
the free parameters of the theory have the values ($\omega$,
$\lambda_1$, $N$, $c$, $\beta$, $m_1$)=(1, -1, 60, 0, 0.017, 10)
then the observed quantities in Eq. (\ref{observed}) obtain values
compatible with the current observational data. In fact, the
spectral indices of the scalar and tensor perturbations, along
with the tensor to scalar ratio, obtain the values $n_S=0.965992$,
$n_T=-4.06329\cdot10^{-6}$ and $r=3.2506\cdot10^{-5}$ which are
accepted values according to the recent Planck 2018 collaboration
\cite{Akrami:2018odb}. Furthermore, we mention that the initial
and final value of the scalar field are $\phi_i=0.0184556$ and
$\phi_f=12.948$ which indicates an increase in the scalar field.
Lastly, we note that the slow-roll indices obtain the values
$\epsilon_1=2.03164\cdot10^{-6}$, $\epsilon_4=6\cdot10^{-29}$,
$\epsilon_5=6\cdot10^{-18}$ and $\epsilon_6=7\cdot10^{-18}$ which
are extremely small.
\begin{figure}[h!]
\centering
\includegraphics[width=17pc]{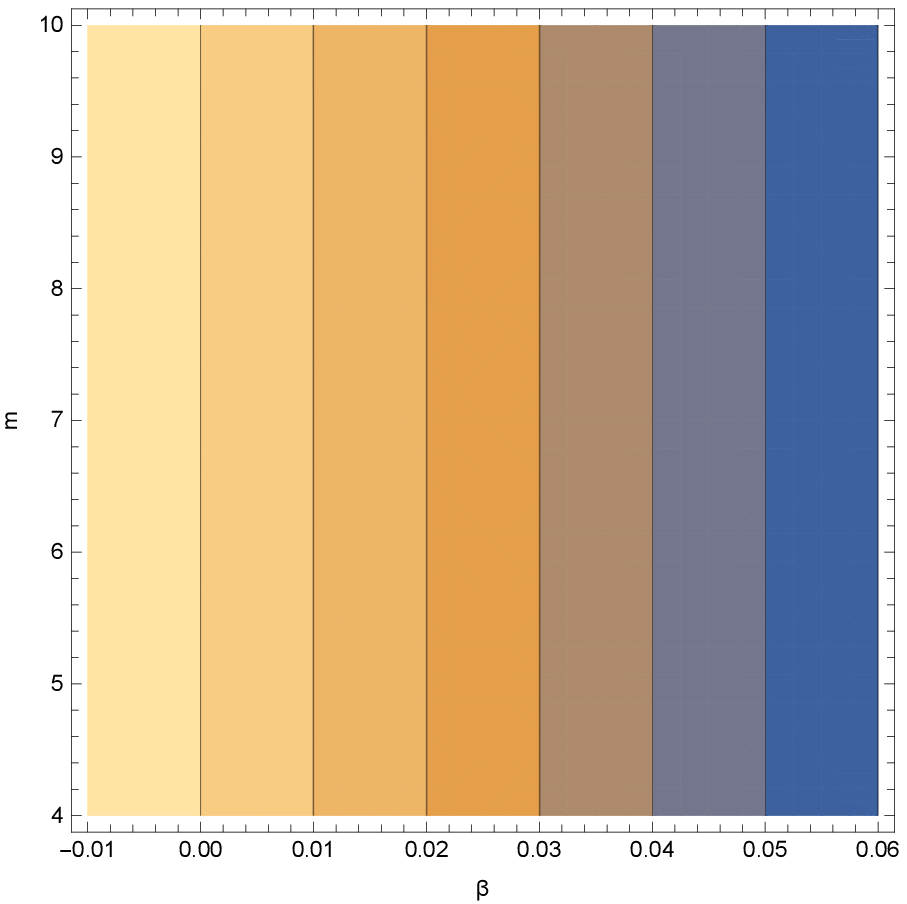}
\includegraphics[width=2pc]{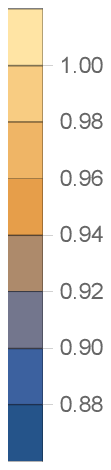}
\includegraphics[width=17pc]{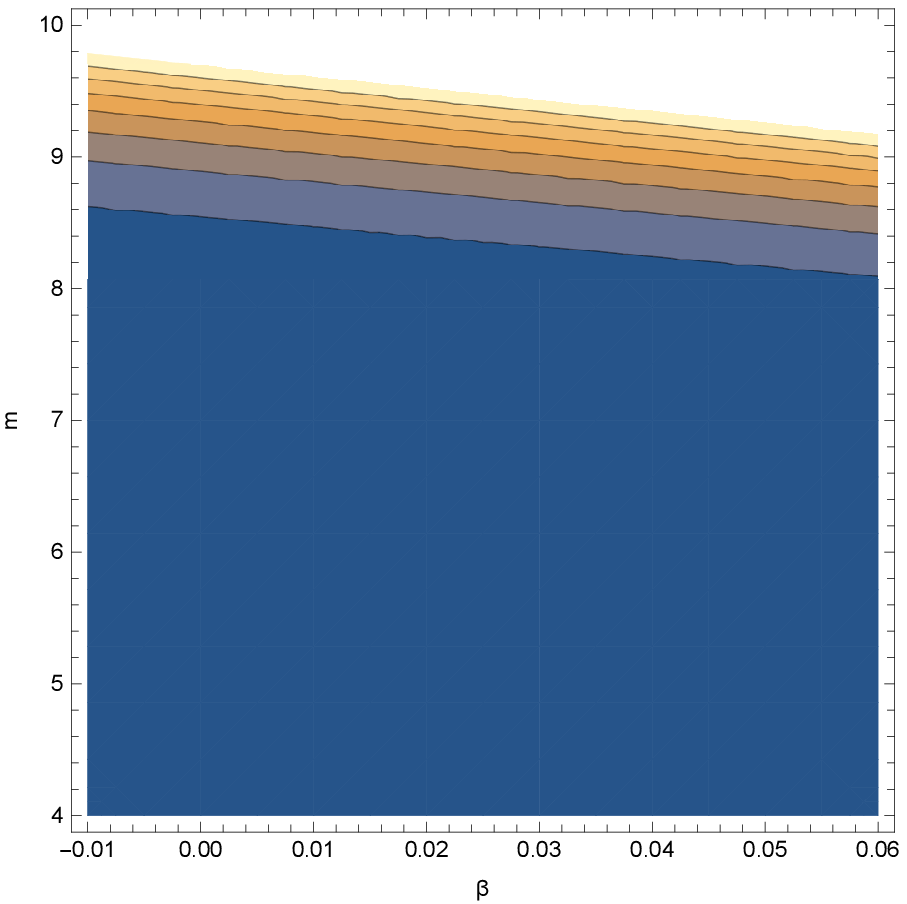}
\includegraphics[width=3pc]{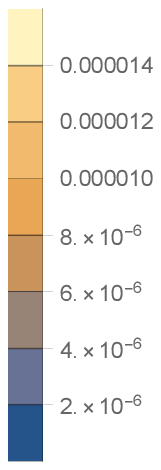}
\caption{Contour plots of the spectral index of primordial
curvature perturbations (right) and the tensor-to-scalar ratio
(left) depending on parameters $\beta$ and $m$, ranging from
[0.01,0.09] and [4,10] respectively. Concerning the spectral
index, it is clear that the dominant parameter which defines its
value is the constant-roll parameter and in fact there exists a
very narrow area of acceptance which ranges approximately from
0.015 to 0.02, exactly where the value in our example resides.}
\label{plot1}
\end{figure}

\begin{figure}[h!]
\centering
\includegraphics[width=17pc]{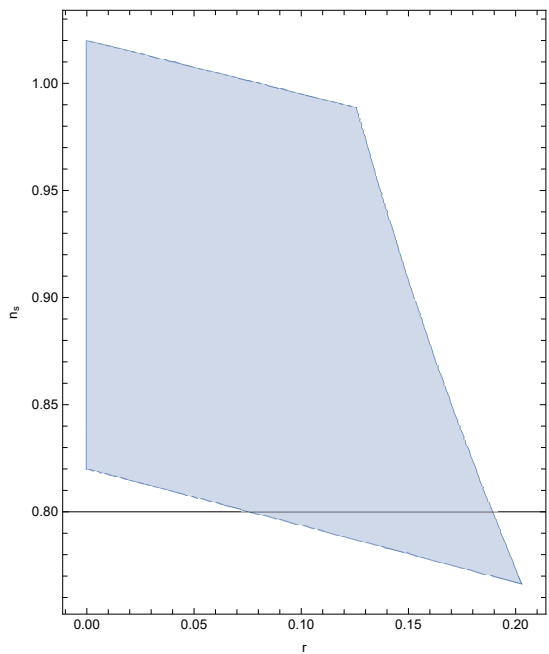}
\caption{Parametric plot of the tensor-to-scalar ratio (x axis)
and the spectral index of scalar perturbations (y axis) depending
on parameters $\beta$ and $m$, ranging from [-0.01, 0.09] and
[4,26] respectively. Even in this case, it is clear that there
exists a narrow area of acceptance for this set parameters due to
the  rage of compatible with the observations values of the
spectral index $n_S$.} \label{plot2}
\end{figure}
Moreover, we make also predictions for the amount of
non-Gaussianities in the primordial power spectrum of the
curvature perturbations. From equations (\ref{NL}), the expected
value of $f_{NL}^{eq}$, for the exact same set of parameters we
used to obtain the viability of the model with the Planck data, is
$f_{NL}^{eq}=0.0910216$ which is also an accepted value and may
explain why non Gaussianities have yet to be observed. Finally,
the parameters used to derive such values are equal to
$\delta_\xi=-10^{-18}$, $\epsilon_s=2.0316\cdot10^{-6}$ and
$\eta=0.21844$ which means that one one of them is in fact
dominant. These results imply that $\epsilon_s=\epsilon_1$.

At this point, it is also worth mentioning that the observed
quantities $n_S$ and $r$ experience different changes when the
values of the free parameters alter. For instance, the
constant-roll parameter $\beta$ is the only one which affects the
spectral index of scalar perturbations while the exponent $m$ of
the coupling scalar function along with the constant-roll
parameter affect the tensor-to-scalar ratio, with the first being
more decisive factor. This can easily be observed in Fig.
\ref{plot1} where one sees that the spectral index of scalar
perturbations is depicted by a simple plot resembling vertical
lines. In addition, while the term $f_{NL}^{eq}$ is independent of
parameter $\lambda$, it can be enhanced by decreasing the exponent
$m$ but such a change leads to a subsequent decrease in the
tensor-to-scalar ratio. For instance, choosing $m=1.5$ leads to
$f_{NL}^{eq}=1.22875$, $n_S=0.966$ and the effective value of the
tensor-to-scalar ratio is 0, since numerically speaking,
$r\sim\mathcal{O}(10^{-102})$. Further information for the
behavior of the spectral index and of the tensor-to-scalar ratio
can be found in Fig. \ref{plot2}, where we present the parametric
plot of the tensor-to-scalar ratio (x axis) and of the spectral
index of scalar perturbations (y axis) depending on parameters
$\beta$ and $m$, ranging from [-0.01, 0.09] and [4,26]
respectively.

Another comment that should be briefly discussed here is the form
of the scalar potential. From the continuity equation, it becomes
apparent that (\ref{potA}) is quite complex, however this is not
true. Since $\xi'\mathcal{G}$ is many orders lesser than $V'$,
something that will be shown shortly, it can be discarded from the
equations of motion leaving us with a scalar potential which is
only exponential, meaning that for $24\xi'H^4\ll V'$, one obtains
the scalar potential,
\begin{equation}
\centering
\label{VA1}
V(\phi)=V_1e^{\frac{\left(\beta ^2+2 \beta -3\right) \kappa ^2 \phi ^2 \omega }{6 (m_1-1)}}\, ,
\end{equation}
which is a quite simple case. This applies obviously to the
slow-roll case studied in Ref. \cite{Odintsov:2020sqy}.

Finally, we examine the validity of the approximations which were
made during this model at the first horizon crossing. Firstly, the
slow-roll approximations in Eq. (\ref{approx}) hold true since
$\dot H\sim\mathcal{O}(10^{-7})$ compared to
$H^2\sim\mathcal{O}(10^{-1})$ and
$\frac{1}{2}\omega\dot\phi^2\sim\mathcal{O}(10^{-7})$ compared to
$V\sim\mathcal{O}(10^{1})$ are negligible. Also, the terms which
were omitted in equations (\ref{motion1}) and (\ref{motion2}) are
of of the order (in reduced Planck units) $24\dot\xi
H^3\sim\mathcal{O}(10^{-17})$ while $16\dot\xi H\dot
H\sim\mathcal{O}(10^{-23})$, which explains why these terms,
compared to the scalar potential and the kinetic term, can be
neglected. Lastly, $V'\sim\mathcal{O}(10^{-3})$ whereas
$\xi'\mathcal{G}\sim\mathcal{O}(10^{-15})$ which explains why the
second form of the scalar potential presented in Eq. (\ref{VA1})
is equivalent to that of (\ref{potA}).

\subsection{Model II: Advanced Exponential Model}

Let us now assume that the coupling scalar function has the
following form,
\begin{equation}
\centering
\label{xi2}
\xi(\phi)=\kappa\lambda_2\int^{\kappa\phi}{e^{\gamma_2 x^{m_2}}dx},\,
\end{equation}
where $x$ is an auxiliary integration variable. This may seem like
a strange choice, but it can be justified due to the simple form
of the ratio $\xi'/\xi''$ which appears in our calculations, as
\begin{equation}
\centering
\xi''=m_2\gamma_2\frac{(\kappa\phi)^{m_2}}{\phi}\xi'.\,
\end{equation}
In order to find the expression of the scalar potential, we must
make use of Eq. (\ref{motion6}). However, the differential
equation is not so easy to solve. To do so, we must make an
additional approximation which is reasonable and it realized in
the following equation,
\begin{equation}
\centering \label{Vdifeq2}
V'+\kappa^2\omega(1-\beta)\left(1+\frac{\beta}{3}\right)\frac{\xi'}{\xi''}V\simeq0.\,
\end{equation}
Using this differential equation, the resulting scalar potential
is,
\begin{equation}
\centering
\label{potB}
V(\phi)=V_2exp(\alpha_2(\kappa\phi)^{2-m_2}),\,
\end{equation}
where here, $\alpha_2=\frac{\omega(\beta^2+2\beta-3)}{3\gamma_2
m_2(2-m_2)}$ and $V_2$ the integration constant with mass
dimensions [m]$^{4}$. Continuing, the resulting expressions for
several terms of interest are shown below,
\begin{equation}
\centering
\label{dxiB}
\delta_\xi\simeq\frac{(1-\beta ) \kappa\phi  \lambda_2  \kappa^4  V(\phi ) (\kappa  \phi )^{-m_2} e^{\gamma_2  (\kappa  \phi )^{m_2}}}{3 \gamma_2  m_2},\,
\end{equation}
\begin{equation}
\centering
\label{esB}
\epsilon_s\simeq\frac{(1-\beta )  (\kappa  \phi )^{1-2 m_2} \left(3 (1-\beta)\kappa \phi  \omega -8 \gamma_2 \lambda_2  m_2\kappa^4 V(\phi ) (\kappa  \phi )^{m_2} e^{\gamma_2  (\kappa  \phi )^{m_2}}\right)}{6 \gamma_2 ^2 m_2^2},\,
\end{equation}
\begin{equation}
\centering
\label{index1B}
\epsilon_1\simeq\frac{\omega}{2}\left(\frac{1-\beta}{m_2\gamma_2}\right)(\kappa\phi)^{2(1-m_2)},\,
\end{equation}
\begin{equation}
\centering
\label{index2B}
\epsilon_2=\beta,\,
\end{equation}
\begin{equation}
\centering
\label{index3B}
\epsilon_3=0,\,
\end{equation}
\begin{equation}
\centering
\label{index5B}
\epsilon_5\simeq\frac{4 (\beta -1) \kappa\phi\lambda_2  \kappa^4  V(\phi ) e^{\gamma_2  (\kappa  \phi )^{m_2}}}{3 \gamma_2  m_2 (\kappa  \phi )^{m_2}+8 (\beta -1) \kappa \phi \lambda_2  \kappa^4  V(\phi ) e^{\gamma_2  (\kappa  \phi )^{m_2}}}\,,
\end{equation}
\begin{equation}
\label{index6B}
\centering
\epsilon_6\simeq-\frac{4 (\beta -1)^2  \lambda_2    (\kappa  \phi )^{1-m_2} e^{\gamma_2  (\kappa  \phi )^{m_2}} \left(m_2\kappa^4 V(\phi ) \left(\gamma_2  (\kappa  \phi )^{m_2}-1\right)+\kappa\phi \kappa^3 V'(\phi )+\kappa^4V(\phi )\right)}{\gamma_2 m_2 \left(3 \gamma_2  m_2 (\kappa  \phi )^{m_2}+8 (\beta -1) \kappa \phi \lambda_2  \kappa^4 V(\phi ) e^{\gamma_2  (\kappa  \phi )^{m_2}}\right)}.\,
\end{equation}
Similar to the previous model, index $\epsilon_4$ was omitted due
to its intricate form. Finally, as was the case with the previous
model, we present the initial and final value of the scalar field,
\begin{equation}
\centering
\label{scalarfB}
\phi_f=\frac{1}{\kappa}\left(\sqrt{\frac{\omega }{2}} \left| \frac{1-\beta }{m_2 \gamma_2 }\right|\right)^{\frac{1}{m_2-1}},\,
\end{equation}
\begin{equation}
\centering
\label{scalariB}
\phi_i=\frac{1}{\kappa}\left((\kappa\phi_f)^{m_2}-\frac{N(1-\beta)}{\gamma_2}\right)^{\frac{1}{m_2}}.\,
\end{equation}
Assuming that in Planck Units, ($\omega$, $\lambda_2$, $N$, $V_2$,
$\beta$, $m_2$, $\gamma$)=(1, 100, 60, 1, 0.017, 3, -1) then the
resulting spectral index of primordial curvature perturbations and
the tensor-to-scalar ratio are compatible with the observations,
as $n_S=0.965059$ and $r=0.003732$ are acceptable values.
Furthermore, we mention that the unobserved spectral index of
tensor perturbations obtains the value $n_T=-0.000466$ and for the
scalar field, $\phi_i=3.89501$ and $\phi_f=0.481347$ which shows
that the scalar field decreases  with time. Finally, when it comes
to the slow-roll indices, the majority of them have extremely
small values as $\epsilon_1=0.00023$,
$\epsilon_4=4.17\cdot10^{-47}$ and
$\epsilon_5=5.7\cdot10^{-26}=\epsilon_6$.
\begin{figure}[h!]
\centering
\includegraphics[width=17pc]{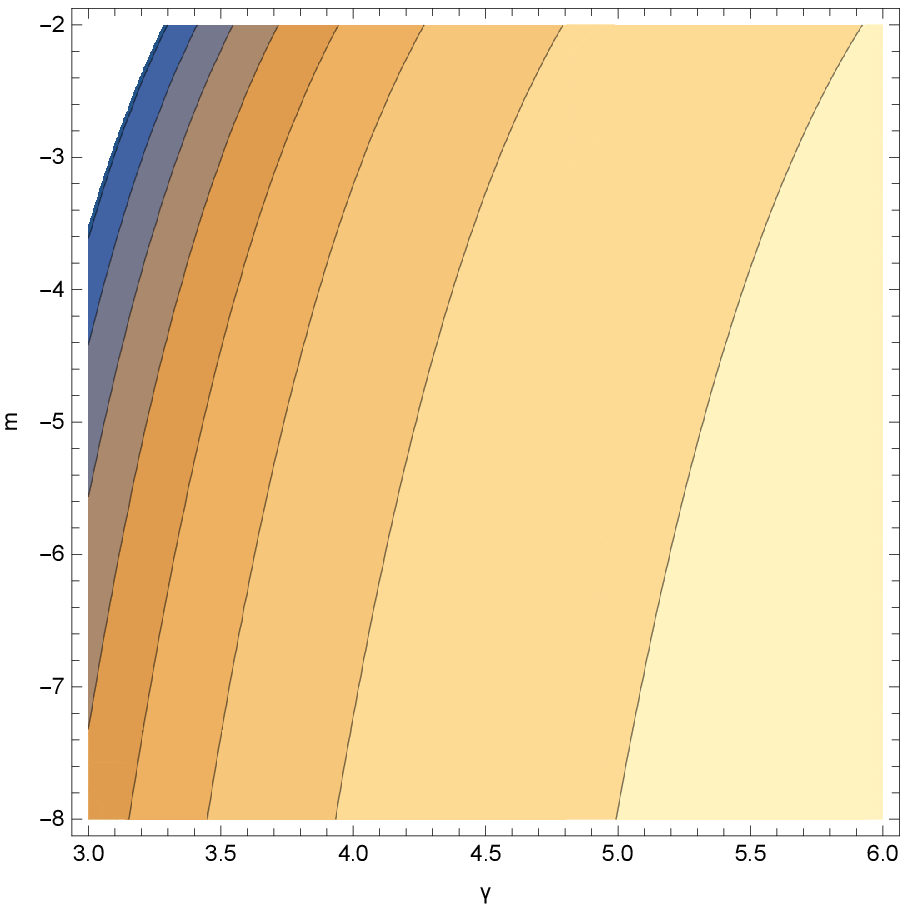}
\includegraphics[width=3pc]{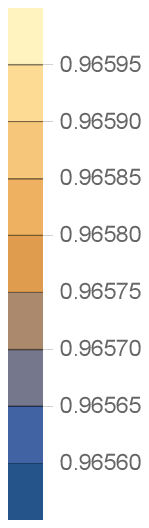}
\includegraphics[width=17pc]{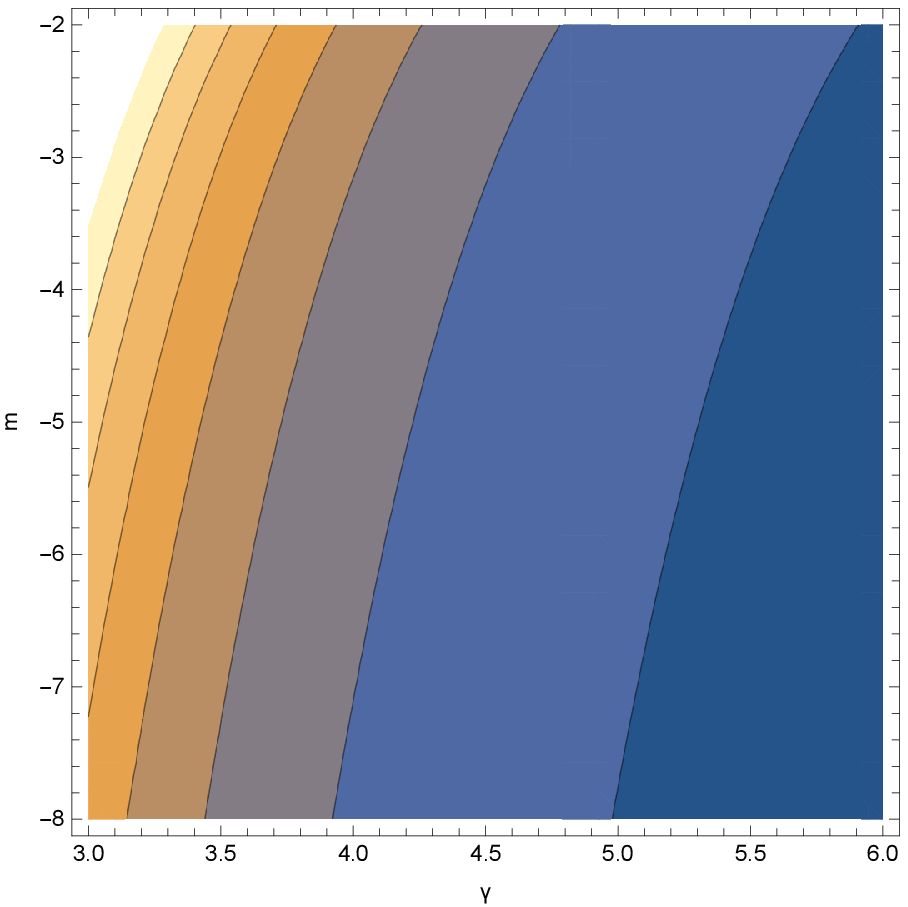}
\includegraphics[width=3pc]{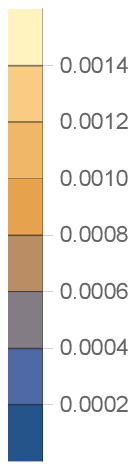}
\caption{Contour plots of the spectral index of primordial
curvature perturbations (left) and the tensor-to-scalar ratio
(right) depending on parameters $m$ and $\gamma$ ranging from
[3,6] and [-8,-2] respectively. It can be inferred that both
parameters influence their values but the spectral index changes
with a lesser rate.} \label{plot3}
\end{figure}
Concerning the non-Gaussianities issue, it turns out that the term
$f_{NL}^{eq}$ obtains the value $f_{NL}^{eq}=0.009598$ which is
obviously small.  Moreover, $\delta_\xi=-1.43\cdot10^{-26}$,
$\epsilon_s=0.00023$ and $\eta=0.02218$ In this model, not only
does $\epsilon_s$ coincide with index $\epsilon_1$, but two of the
three parameters for evaluating the non-Gaussianities have non
negligible values.

In this case, the exponent $m$ affects both the spectral index of
scalar perturbations and the tensor-to-scalar ratio. The same
applies to the constant-roll parameter $\beta$ but in this case,
only the spectral index experiences a significant change. Lastly,
$\gamma_2$ alters both values and as a matter of fact in not so
significant rate. Decreasing $\gamma_2$ to the value -10 alters
the fourth decimal in each magnitude. In contrast, if $\beta$ was
to obtain the value 0.015, which is a really small change, the
spectral index takes a non-compatible value with the observations,
which indicates the great impact such a change in the parameters
has. The dependence on $\beta$ and $\gamma_2$ can be viewed in
Fig. \ref{plot3}. Finally, the exponent may vary in the range
[3,14] and the only change which the observed quantities shall
experience is in the fourth and fifth decimal, with the
tensor-to-scalar ratio experiencing a decrease.

Let us proceed with the dynamics of the system in terms of
altering certain free parameters. It turns out that many
parameters leave the results unaltered, while others play a
significant role. For instance, changing the constant-roll
parameter $\beta$ to 0.015, while it affects greatly the spectral
index of primordial curvature perturbations, it does not alter the
order of magnitude of $f_{NL}^{eq}$, just changes the numerical
value in the same order. Even if it changes to, lets say
$\beta=0.6$, the results are the same. In contrast, the exponent
$m$ affects significantly the parameters, since an increase in the
exponent, leads to a decrease in the tensor-to-scalar ratio but
also enhances the non-Gaussianities and even by a lot. As it was
shown in the slow-roll case \cite{Odintsov:2020sqy}, the exponent
$m_2$ is free to take values at least up to 120. The term
$f_{NL}^{eq}$ is not an interesting case to examine since as
mentioned in the previous section, it can be evaluated by using
the exact same auxiliary parameters as the spectral indices of
scalar, tensor perturbations and the tensor-to-scalar ratio, so it
is expected to be dependent on the same free parameters as those
quantities. However, we mention that changing the exponent, for
instance to $m=100$ leads to viable results and also increases the
non linear term only by one order, meaning that
$f_{NL}^{eq}=0.0137485$.

Finally, we discuss the approximations made throughout the
equations of motion. When it comes to the slow-roll
approximations, we note that $\dot H\sim\mathcal{O}(10^{-5})$
while $H^2\sim\mathcal{O}(10^{-1})$ and similarly,
$\frac{1}{2}\omega\dot\phi^2\sim\mathcal{O}(10^{-5})$ and
$V\sim\mathcal{O}(10^{-1})$ which shows that the approximations in
fact do apply. In addition, the following terms are $24\dot\xi
H^3\sim\mathcal{O}(10^{-25})$, $16\dot\xi H\dot
H\sim\mathcal{O}(10^{-29})$ which justifies why they were
neglected in equations (\ref{motion1}) and (\ref{motion2}).
Furthermore, for equation (\ref{motion3}),
$V'\sim\mathcal{O}(10^{-2})$ in contrast to
$24\xi'H^4\sim\mathcal{O}(10^{-23})$ so it is reasonable why the
latter was neglected.

As a last comment, we mention that the dominant parameters which
affect the results are mainly the constant-roll parameter $\beta$
and parameters $\gamma_2$ and $m_2$ while $\lambda_2$ seems to not
cause any change to the results along with $V_2$.

\subsection{Comparison Between Slow-roll And Constant-Roll}

In this case, we shall work with the previous model but implement
a different formalism. Here, we shall set $\beta$ equal to zero,
so visually it will disappear from all the previous equations but
in reality, for the scalar field we assume that in addition to the
slow-roll approximations (\ref{approx}), the approximation
$\ddot\phi\ll\dot\phi H$ holds. This will lead to a change with
the new set of equations being,
\begin{equation}
\centering
\label{xiC}
\xi(\phi)=\kappa\lambda_3\int^{\kappa\phi}{e^{\gamma_3 x^m}dx},\,
\end{equation}
\begin{equation}
\centering
\dot\phi\simeq H\frac{\xi'}{\xi''},\,
\end{equation}
\begin{equation}
\centering
H^2\simeq\frac{\kappa^2V}{3},\,
\end{equation}
\begin{equation}
\centering
\dot H\simeq -\frac{H^2}{2}\kappa^2\omega\left(\frac{\xi'}{\xi''}\right)^2,\,
\end{equation}
\begin{equation}
\centering
V'+\omega\kappa^2\frac{\xi'}{\xi''}V\simeq0.\,
\end{equation}
This model was studied thoroughly in \cite{Odintsov:2020sqy} and
it is capable of producing viable results. Before we proceed with
the results however, it is worth mentioning the changes to which
the scalar potential and the slow-roll indices will be subjected
to. From the previous set of equations, the resulting scalar
potential is,
\begin{equation}
\centering
\label{potC}
V(\phi)=V_3e^{-\frac{\omega(\kappa \phi) ^{2-m_3} }{\gamma_3  (2-m_3) m_3}},\,
\end{equation}
which as expected is the same as before, (\ref{potB}) with
$\beta=0$. Similarly,
\begin{equation}
\centering
\label{dxiC}
\delta_\xi\simeq\frac{\kappa \phi \lambda_3  \kappa^4  V(\phi ) (\kappa  \phi )^{-m_3} e^{\gamma_3  (\kappa  \phi )^{m_3}}}{3 \gamma_3  m_3},\,
\end{equation}
\begin{equation}
\centering
\label{esC}
\epsilon_s\simeq\frac{  (\kappa  \phi )^{1-2 m_3} \left(3 \kappa\phi  \omega -8 \gamma_3 \lambda_3  m_3\kappa^4 V(\phi ) (\kappa  \phi )^{m_3} e^{\gamma_3  (\kappa  \phi )^{m_3}}\right)}{6 \gamma_3 ^2 m_3^2},\,
\end{equation}
\begin{equation}
\centering
\label{index1C}
\epsilon_1\simeq\frac{\omega  (\kappa  \phi )^{2(1-m_3)}}{2 (\gamma_3 m_3)^2},\,
\end{equation}
\begin{equation}
\centering
\label{index2C}
\epsilon_2\simeq-\frac{(\kappa  \phi )^{-2 m_3} \left(\omega(\kappa \phi )^2 +2m_3(m_3-1) \gamma_3 (\kappa  \phi )^{m_3}\right)}{2 \gamma_3 ^2 m_3^2},\,
\end{equation}
\begin{equation}
\centering
\label{index3C}
\epsilon_3=0,\,
\end{equation}
\begin{equation}
\centering
\label{index5C}
\epsilon_5\simeq\frac{4 \kappa \phi \lambda_3  \kappa^4 V(\phi ) e^{\gamma_3  (\kappa  \phi )^{m_3}}}{8 \kappa\phi \lambda_3  \kappa^4  V(\phi ) e^{\gamma_3  (\kappa  \phi )^{m_3}}-3 \gamma_3  m_3 (\kappa  \phi )^{m_3}},\,
\end{equation}
\begin{equation}
\centering
\label{index6C}
\epsilon_6\simeq-\frac{4 \kappa \phi \lambda_3    (\kappa  \phi )^{-m_3} e^{\gamma_3  (\kappa  \phi )^{m_3}} \left(m_3 \kappa^4V(\phi ) \left(\gamma_3  (\kappa  \phi )^{m_3}-1\right)+\kappa\phi \kappa^3 V'(\phi )+\kappa^4V(\phi )\right)}{\gamma_3  m_3 \left(3 \gamma_3  m_3 (\kappa  \phi )^{m_3}-8 \kappa \phi \lambda_3  \kappa^4  V(\phi ) e^{\gamma_3  (\kappa  \phi )^{m_3}}\right)}.\,
\end{equation}
Similarly, apart from $\epsilon_2$, all the indices coincide with
those previously for $\beta=0$, as expected. Following, the exact
same steps, the initial and final value of the scalar field read,
\begin{equation}
\centering
\label{scalarfC}
\phi_f=\frac{1}{\kappa}\left(\sqrt{\frac{\omega}{2}}\left|\frac{1}{\gamma_3 m_3}\right|\right)^{\frac{1}{(m_3-1)}},\,
\end{equation}
\begin{equation}
\centering
\label{scalariC}
\phi_i=\frac{1}{\kappa}\left((\kappa\phi)^{m_3}-\frac{N}{\gamma_3}\right)^{\frac{1}{m_3}}.\,
\end{equation}
Assuming that in Planck Units, ($\omega$, $\lambda_3$, $N$, $V_3$,
$m_3$, $\gamma_3$)=(1, 1, 60, 1, 20, -0.001) then from Eq
(\ref{observed}), we obtain the values $n_S=0.968331$,
$n_T=-2.08667\cdot10^{-6}$ and $r=1.6693\cdot10^{-5}$ which are
compatible results with the observations results. Moreover,
$\phi_i=1.7335$ and $\phi_f=1.20642$ in Planck Units, which shows
a decrease with time. And finally, $\epsilon_1=1.04\cdot10^{-6}$,
$\epsilon_4=7\cdot10^{-52}$ and $\epsilon_5=10^{-29}=\epsilon_6$
which indicates that the slow-roll conditions indeed apply.

The main aim of the analysis performed in this subsection however,
was to evaluate and predict the amount of non-Gaussianities in the
power spectrum. The above set of parameters leads to the value
$f_{NL}^{eq}=0.013196$. Similarly, $\delta_\xi=-4\cdot10^{-30}$,
$\epsilon_s=\epsilon_1$ and $\eta=0.03164$. A quick comparison
between the models corresponding to the slow-roll and
constant-roll case, indicate that in the constant-roll case the
value of $f_{NL}^{eq}$ decreases, since viability can be achieved
for smaller values of the exponent $m$. Thus the main difference
between the two phenomenologies is the set of values for the free
parameters that can achieve both viability for the observed the
spectral index of primordial curvature perturbations and the
tensor-to-scalar ratio. In conclusion, both the slow-roll and the
constant-roll condition of this particular model are more than
capable of describing a viable phenomenology and in fact are able
to predict the same amount of non-Gaussianities, and remarkably in
the constant-roll case, slightly smaller amount of
non-Gaussianities.

\section{Phenomenology by Imposing the Condition $\kappa\xi'/\xi''\ll 1$}

In this section we shall assume that the following condition holds
true $\kappa\xi'/\xi''\ll 1$, and we shall examine the
phenomenological implications for an appropriately chosen model.
Thus the differential equation that connects the scalar potential
and the scalar coupling function takes the form,
\begin{equation}
\centering
\label{Vdifswamp}
V'+\frac{8}{3}\kappa^4\xi'V^2\simeq0.\,
\end{equation}
This is a simple ordinary differential equation which has the
following solution,
\begin{equation}
\centering
V(\phi)=\frac{1}{\frac{8}{3}\kappa^4\xi(\phi)-\Lambda},\,
\end{equation}
where $\Lambda$ is an integration constant with mass dimensions
[m]$^{-4}$. By appropriately choosing the Gauss-Bonnet coupling
scalar function, specifies immediately the scalar potential. In
this model, let us assume that the coupling function is defined
as,
\begin{equation}
\centering
\label{xi6}
\xi(\phi)=\lambda_4 Erf(\gamma_4\kappa\phi).\,
\end{equation}
This is a model which was also studied in our previous work
\cite{Odintsov:2020sqy}. It is an appropriate function since,
\begin{equation}
\centering
\xi''=-2(\gamma_4\kappa)^2\phi\xi',\,
\end{equation}
thus the ratio $\xi'/\xi''$ is greatly simplified. In addition,
the corresponding slow-roll indices are written as,
\begin{equation}
\centering
\label{dxiF}
\delta_\xi\simeq-\frac{(1-\beta)\lambda_4  \kappa^4V(\phi ) e^{-(\gamma_4  \kappa \phi) ^2}}{3 \sqrt{\pi } \gamma_4 \kappa \phi },\,
\end{equation}
\begin{equation}
\centering
\label{esF}
\epsilon_s\simeq\frac{(1-\beta ) \left(32 \gamma_4 ^3 \kappa\phi \lambda_4  \kappa^4  V(\phi ) e^{-(\gamma_4 \kappa \phi) ^2}+3\sqrt{\pi}\omega (1-\beta )  \right)}{6\sqrt{\pi}(2 \gamma_4 ^2 \kappa  \phi) ^2},\,
\end{equation}
\begin{equation}
\centering
\label{index1F}
\epsilon_1\simeq\frac{\omega}{2}\left(\frac{1-\beta}{2\gamma_4^2\kappa\phi}\right)^2,\,
\end{equation}
\begin{equation}
\centering
\label{index2F}
\epsilon_2=\beta,\,
\end{equation}
\begin{equation}
\centering
\label{index3F}
\epsilon_3=0,\,
\end{equation}
\begin{equation}
\centering
\label{index5F}
\epsilon_5\simeq\frac{4 (1-\beta) \lambda_4\kappa^4  V(\phi )}{8 (1-\beta) \lambda_4  \kappa^4V(\phi )+3 \sqrt{\pi } \gamma_4\kappa  \phi  e^{(\gamma_4 \kappa \phi )^2}},\,
\end{equation}
\begin{equation}
\centering
\label{index6F}
\epsilon_6\simeq\frac{2 (1-\beta)^2 \kappa  \lambda_4  \left(-\kappa\phi \kappa^3 V'(\phi )+(2 (\gamma_4\kappa \phi) ^2+1)\kappa^4 V(\phi ))\right)}{(\gamma_4 \kappa \phi)^2 \left(8 (1-\beta ) \lambda_4 \kappa^4 V(\phi )+3 \sqrt{\pi } \gamma_4  \kappa\phi  e^{(\gamma_4 \kappa  \phi)^2}\right)}.\,
\end{equation}
Finally, we mention that the values of the scalar field during the
initial and final moment of inflation are,
\begin{equation}
\centering
\label{scalarfF}
\phi_f=-\sqrt{\frac{\omega}{2}}\frac{|1-\beta|}{2\gamma_4^2\kappa},\,
\end{equation}
\begin{equation}
\centering
\label{scalariF}
\phi_i=\frac{1}{2\gamma_4^2\kappa}\sqrt{\frac{8N\gamma_4^2+\omega(1+\beta^2-2\beta)}{2}}.\,
\end{equation}
Assuming that in Planck Units, ($\omega$, $\lambda_4$, $N$,
$\Lambda$, $\beta$, $\gamma_4$)=(1, $10^4$, 60, 0, 0.013, 1) then
the resulting values for the spectral index of primordial
curvature perturbations and the tensor-to-scalar ratio are
$n_S=0.965829$ and $r=0.0324065$, which are are both compatible
results with the Planck 2018 data \cite{Akrami:2018odb}.
Furthermore, the spectral index of tensor perturbations is
$n_T=-0.00405904$ and the values of the scalar field are
$\phi_i=7.75382$ and $\phi_f=-0.348957$ which indicates a decrease
in the scalar potential. And finally, when it comes to the
slow-roll indices, $\epsilon_1=0.00202$,
$\epsilon_4=1.13\cdot10^{-52}$,
$\epsilon_5=2.7\cdot10^{-28}=\epsilon_6$. The effective value of
the last three is obviously zero.

In addition, the predicted values for the non-Gaussianities are
also compatible results. We mention that the equilateral non
linear term obtains the value $f_{NL}^{eq}=0.009934$ which is
quite a small value. Also, $\delta_\xi=-6.95\cdot10^{-29}$,
$\epsilon_s=\epsilon_1$, $\eta=0.0164167$. The $f_{NL}^{eq}$ term
can obtain a greater value by decreasing $\gamma_4$, but such
decrease leads to a subsequent increase in the tensor-to-scalar
ratio so it must be made with care. Choosing $\gamma_4=0.8$ leads
to $f_{NL}^{eq}=0.011662$ while producing also viable spectral
indices and tensor-to-scalar ratio. Here, $\gamma_4$ affects the
following quantities, the tensor-to-scalar ratio, the term
$f_{NL}^{eq}$ and the spectral index of scalar perturbations,  but
mainly the first two, while the latter are affected greatly by the
constant-roll parameter $\beta$.

Lastly, we examine the validity of our approximations. Concerning
the slow-roll approximations, we note that $\dot
H\sim\mathcal{O}(10^{-8})$ compared to
$H^2\sim\mathcal{O}(10^{-5})$ similarly
$\frac{1}{2}\omega\dot\phi^2\sim\mathcal{O}(10^{-8})$ in contrast
to $V\sim\mathcal{O}(10^{-5})$. Indeed, the approximations in
(\ref{approx}) are valid. Moreover, the string terms in the
equations of motion (\ref{motion1}) and ({\ref{motion2}) are
$24\dot\xi H^3\sim\mathcal{O}(10^{-32})$ and $16\dot\xi H\dot
H\sim\mathcal{O}(10^{-35})$ which justifies the reason the were
neglected. Also, the ratio $\xi'/\xi''$ is of order
$\mathcal{O}(10^{-3})$. The term $V'$ is of same order as
$\xi'V^2$.

This set of values for the free parameters of the theory is
interesting due to the fact that selecting $\Lambda=0$ implies
that,
\begin{equation}
\centering
V(\phi)=\frac{3}{8\kappa^4\xi(\phi)},\,
\end{equation}
which is an interesting relation between the scalar functions of the model.

\section{Conclusions}

In this work we investigated the quantitative effects of imposing
a constant-roll evolution on the scalar field for a
GW170817-compatible Einstein-Gauss-Bonnet theory. Our focus was on
the inflationary era, and we calculated the slow-roll indices and
the observational quantities of inflation, and we confronted
several models with the observational data coming from the Planck
2018 collaboration. As we demonstrated, the resulting inflationary
phenomenology can be compatible with the latest Planck data, for a
wide range of the free parameters of the theory, and with the
constant-roll condition holding true. In our calculations we
demonstrated that all the assumptions we made were satisfied for
all the models we examined, and for the values of the free
parameters that yield inflationary viability with respect to the
latest Planck data. For all the models we studied, we also
investigated the amount of non-Gaussianities that are predicted
from the models, by calculating the quantity $f_{NL}^{eq}$ in the
equilateral momentum approximation. Interestingly enough, we
demonstrated that the amount of non-Gaussianities is quite small
in the constant-roll case, and also in some cases, where we
compared the slow-roll and constant-roll cases explicitly, we
showed that the quantity $f_{NL}^{eq}$ is even smaller in the
constant-roll case, compared to the slow-roll case. Finally, we
performed an analytic approximation in the differential equation
that connects the scalar field potential and the scalar coupling
function, and we examined the phenomenology of inflation in this
case too. As we evinced, the model can also be compatible with the
Planck 2018 too.

A future study should address the important feature of having the
constraint $c_T^2=1$ holding true after the slow-roll or
constant-roll era, during the reheating era and beyond. In that
case, the differential equation that connects the scalar field
potential $V(\phi)$ and the scalar coupling function $\xi (\phi)$
is not simplified, as it was during the inflationary era, thus one
may use the exact form of the differential equation, and impose
the constraint that it holds true for all the post-inflationary
eras, and that it exactly defines the interconnection of the
scalar potential and of the scalar-Gauss-Bonnet coupling. This
differential equation could be taken as an additional constraint
in the theory, and may affect the reheating era if the
constant-roll assumption is used, like for example in Ref.
\cite{Odintsov:2020sqy}, or even if the slow-roll assumption is
used. The point is that in our previous work
\cite{Odintsov:2020sqy}, and in the present work, we found only an
approximate relation for the scalar potential and the
scalar-Gauss-Bonnet coupling function, however, in the
post-inflationary era, the full differential equation should be
taken into account, and thus, one may have the exact relation
between the two scalar functions, without the need of any
approximation. Thus, one may use this differential equation as an
additional constraint, and solve thus numerically problems of
astrophysical or cosmological interest. This task is in our future
plans and we aim to materialize this in the next years, motivated
by the current astrophysical and cosmological interest on the
gravitational wave speed
\cite{Nair:2019iur,Carson:2020cqb,Giare:2020vss}.

\end{document}